\pdfoutput=1

\documentclass[twocolumn,superscriptaddress,aps,preprintnumbers,amsmath,amssymb,prl,nofootinbib]{revtex4}

\usepackage{epsfig}
\usepackage{amsmath}
\usepackage{amssymb}
\usepackage{subfigure}
\usepackage{cancel}
\usepackage{textcomp}
\usepackage{calc}
\usepackage{graphicx}
\usepackage{dcolumn}
\usepackage{color}
\usepackage{xcolor}
\usepackage{pifont}
\usepackage{slashed}
\usepackage{fdsymbol}

\usepackage{hyperref}
\hypersetup{colorlinks=true,urlcolor=blue,linkcolor=red,citecolor=green!60!black}

\begin{document}


\preprint{CERN-TH-2020-151}

\title{Has NANOGrav found first evidence for cosmic strings?}

\author{Simone Blasi}
\email{blasi@mpi-hd.mpg.de}
\affiliation{Max-Planck-Institut  f{\"u}r  Kernphysik,  69117  Heidelberg,  Germany}

\author{Vedran Brdar}
\email{vbrdar@mpi-hd.mpg.de}
\affiliation{Max-Planck-Institut  f{\"u}r  Kernphysik,  69117  Heidelberg,  Germany}

\author{Kai Schmitz}
\email{kai.schmitz@cern.ch}
\affiliation{Theoretical Physics Department, CERN, 1211 Geneva 23, Switzerland}


\begin{abstract}
The \textit{North American Nanohertz Observatory for Gravitational Waves} (NANOGrav) has recently reported strong evidence for a stochastic common-spectrum process affecting the pulsar timing residuals in its 12.5-year data set.
We demonstrate that this process admits an interpretation in terms of a stochastic gravitational-wave background emitted by a cosmic-string network in the early Universe.
We study stable Nambu--Goto strings in dependence of their tension $G\mu$ and loop size $\alpha$ and show that the entire viable parameter space will be probed by an array of future experiments. 
\end{abstract}


\date{\today}
\maketitle


\noindent\textbf{Introduction\,---\,}%
Many models of new physics beyond the Standard Model predict cosmological phase transitions in the early Universe that lead to the spontaneous breaking of an Abelian symmetry~\cite{Mazumdar:2018dfl}.
An exciting phenomenological consequence of such phase transitions is the generation of a network of cosmic strings~\cite{Kibble:1976sj,Jeannerot:2003qv}, vortexlike topological defects that restore the broken symmetry at their core~\cite{Nielsen:1973cs}.
Cosmic strings can form closed loops that lose energy and shrink via the emission of \textit{gravitational waves} (GWs)~\cite{Vachaspati:1984gt,Auclair:2019wcv}.
Indeed, numerical simulations of cosmic strings based on the Nambu--Goto action~\cite{Ringeval:2005kr,BlancoPillado:2011dq} show that this is the dominant energy loss mechanism of cosmic-string loops, if the underlying broken symmetry corresponds to a local gauge symmetry.
The primordial GW signal from a cosmic-string network, which encodes crucial information on ultraviolet physics far beyond the reach of terrestrial experiments, is therefore a major target of ongoing and upcoming searches for a \textit{stochastic GW background} (SGWB)~\cite{Maggiore:1999vm,Romano:2016dpx,Caprini:2018mtu,Christensen:2018iqi}.


A cosmic-string-induced SGWB is expected to stretch across a vast range of GW frequencies, making it an ideal signal for multifrequency GW astronomy.
At high frequencies in the milli- to kilohertz range, the signal can be searched for in space- and ground-based GW interferometers, while at low frequencies in the nanohertz range, \textit{pulsar timing array} (PTA) experiments are sensitive to the signal.
In this Letter, we shall investigate the latter possibility, a cosmic-string-induced GW signal at nanohertz frequencies, in light of the recent results reported by the \textit{North American Nanohertz Observatory for Gravitational Waves} (NANOGrav) PTA experiment~\cite{McLaughlin:2013ira,Brazier:2019mmu}.


In~\cite{Arzoumanian:2020vkk}, the NANOGrav Collaboration presents its results of a search for an isotropic SGWB based on its 12.5-year data set.
Remarkably enough, this study yields strong evidence for the presence of a stochastic process across the 45 pulsars included in the analysis.
The interpretation of the observed signal in terms of a common-spectrum process is strongly preferred over independent red-noise processes (a Bayesian model comparison yields a $\log_{10}$ Bayes factor larger than $4$); however, a conclusive statement on the physical origin of the signal is currently not yet feasible.
In order to qualify as the detection of a GW signal, the pulsar timing residuals would need to exhibit characteristic angular correlations, which are described by the \textit{Hellings--Downs} (HD) curve~\cite{Hellings:1983fr}, the overlap reduction function for pairs of pulsars in the PTA.
Definite evidence for HD interpulsar spatial correlations is, however, not yet present in the 12.5-year data set; the no-correlations hypothesis is only mildly rejected with a $p$ value at the level of around $5\,\%$.
At the same time, a number of systematic effects might be responsible for the signal or at least contribute to it, such as, \textit{e.g.}, pulsar spin noise~\cite{Lam:2016iie} or solar system effects~\cite{Hobbs:2006cd}.
A clear identification of the signal origin therefore requires further work, in particular, independent analyses and larger data sets.
According to~\cite{Arzoumanian:2020vkk}, several such analyses are currently in preparation, which seem to point to results consistent with those reported by the NANOGrav Collaboration.


\begin{figure*}
\centering
\includegraphics[width=0.450\textwidth]{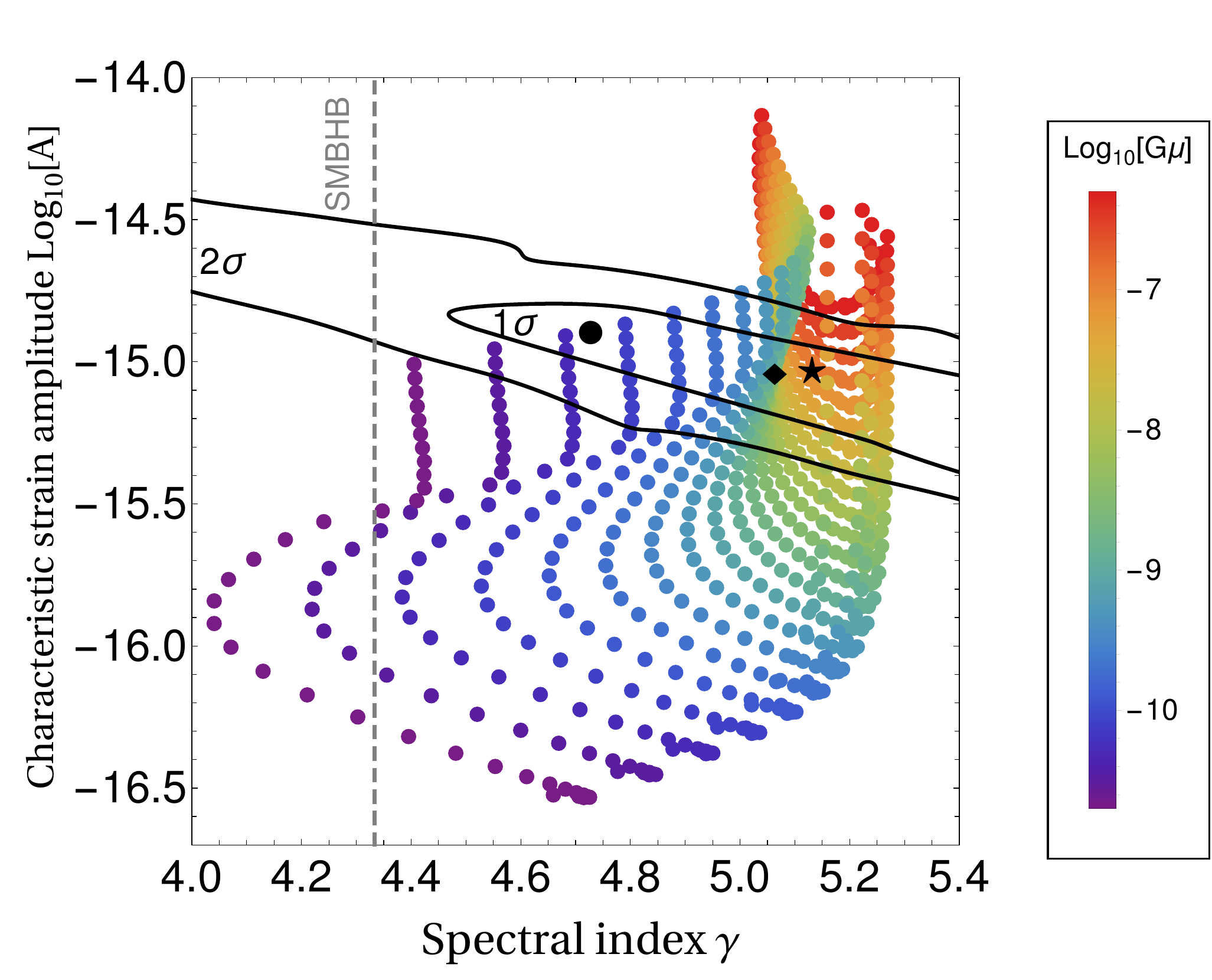}\quad
\includegraphics[width=0.446\textwidth]{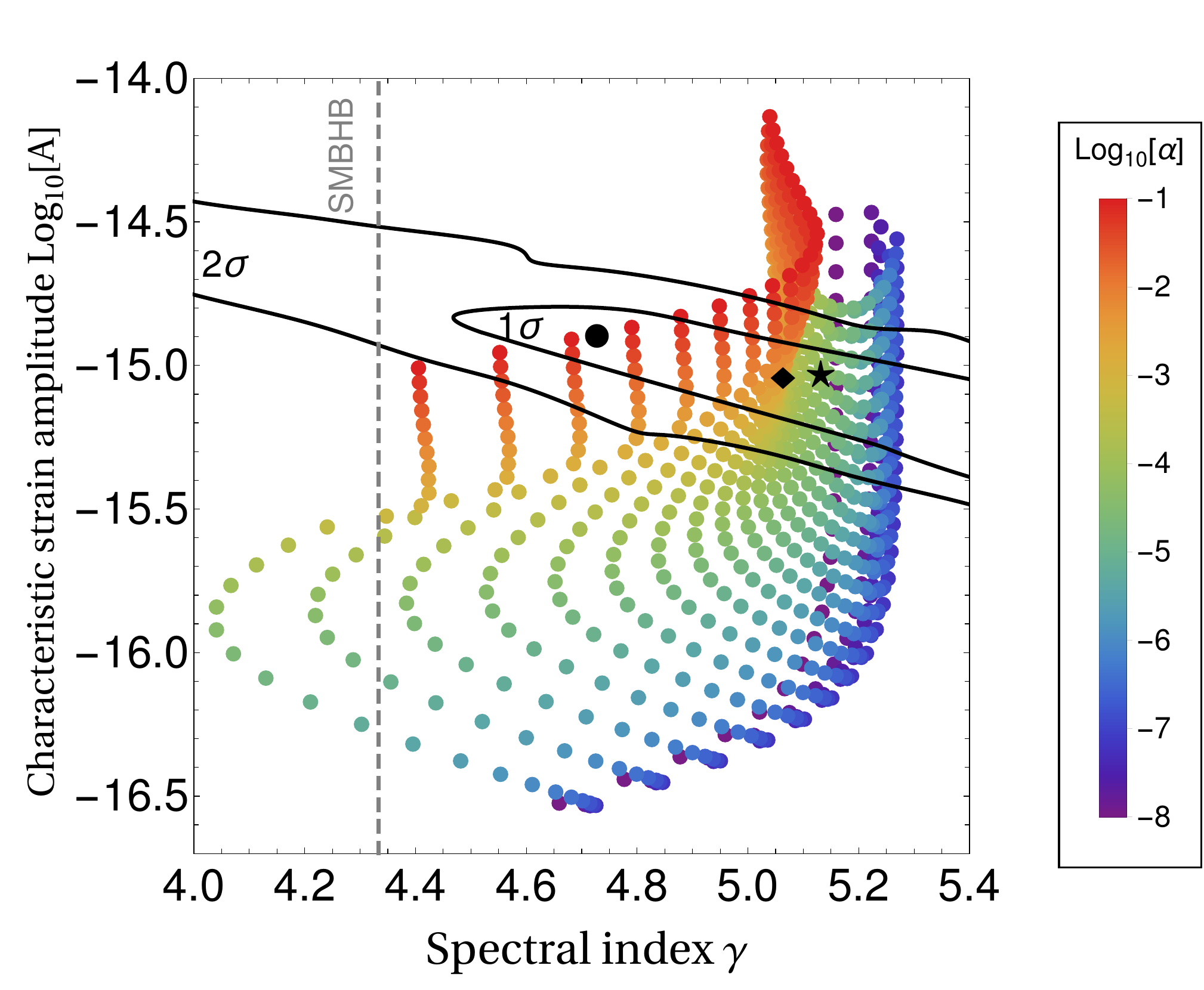}
\caption{Scan over the cosmic-string tension $G\mu$ and loop size $\alpha$ projected onto the $\gamma$\,--\,$A$ plane, where $-\gamma$ represents the spectral index of the pulsar timing-residual cross-power spectrum and $A$ is the characteristic GW strain amplitude at $f = f_{\rm yr}$.
The black contours denote the $1\,\sigma$ and $2\,\sigma$ posteriors in the NANOGrav analysis that allow to describe the observed stochastic process.
Here, we use the contours based on the five lowest frequency bins in the NANOGrav data (see~\cite{Arzoumanian:2020vkk} for details).
The gray vertical line indicates the theoretical prediction for a population of SMBHBs, $\gamma = 13/3$.
The parameter values of the benchmark points ($\smallblackstar$, $\smallblackdiamond$, $\smallblackcircle$) are listed in Tab.~\ref{tab:bp}.
For $\gamma < 5$ ($\gamma > 5$), the GW spectrum is rising (decreasing) as a function of frequency.
In this case, NANOGrav observes GWs at frequencies below (above) the radiation--matter-equality peak in the spectrum.
At the same time, most of the points clustering around $\gamma \simeq 5$ belong to the flat plateau in the spectrum at frequencies above the peak.
}
\label{fig:Agamma}
\end{figure*}


If interpreted in terms of GWs, the NANOGrav signal indicates a GW amplitude at nanohertz frequencies that exceeds previous upper bounds.
This is remarkable and can be traced back to several factors, the most important of which being the choice of a uniform Bayesian prior on the amplitude of pulsar-intrinsic red-noise processes, which shifted signal power to red-noise power in previous PTA analyses~\cite{Hazboun:2020kzd}.
The new NANOGrav results therefore reinvigorate SGWB scenarios that had previously been believed to be severely constrained by PTA bounds. 
In addition to a cosmic-string-induced SGWB~\cite{Siemens:2006yp,Blanco-Pillado:2017rnf,Ringeval:2017eww}, this also includes an astrophysical GW background from merging \textit{supermassive black-hole binaries} (SMBHBs)~\cite{Rajagopal:1994zj,Phinney:2001di,Jaffe:2002rt,Wyithe:2002ep} and a primordial cosmological GW signal from inflation~\cite{Starobinsky:1979ty,Rubakov:1982df,Guzzetti:2016mkm}; see also~\cite{Kobakhidze:2017mru}.
The merger rate of SMBHBs is, however, not known at present; in particular, environmental interactions such as dynamical friction and stellar  scattering are necessary~\cite{Begelman:1980vb} to solve the final-parsec problem and achieve sub-parsec separations in SMBHB systems~\cite{Yu:2001xp,Milosavljevic:2002bn}.
The primordial signal from inflation, on the other hand, is tightly constrained by observations of the \textit{cosmic microwave background} (CMB)~\cite{Ade:2018gkx} and thus requires one to assume a large blue tensor index.
For these reasons, we are going to focus on GWs from cosmic strings in this Letter.
As we are able to show, the simplest model of cosmic Nambu--Goto strings provides a good fit to the NANOGrav signal across large regions of the cosmic-string parameter space. 
These findings promise to open the door to a bright future in GW astronomy.
If NANOGrav should have indeed found first evidence for cosmic strings, future GW experiments will have excellent chances to probe the same signal across more than ten orders of magnitude in GW frequency.
A firm confirmation of the signal will moreover not only drive the field of GW astronomy in the coming years and decades, it will also have a profound impact on particle physics and our understanding of the early Universe.


\smallskip\noindent\textbf{Model\,---\,}%
In this Letter, we shall consider the GW signal from a network of cosmic strings that follows from the cosmological breaking of a generic $U(1)$ gauge symmetry after inflation.
A prime example of such a symmetry, which has lately received a lot of attention~\cite{Buchmuller:2013lra,Dror:2019syi,Buchmuller:2019gfy,Blasi:2020wpy,Fornal:2020esl}, would be $U(1)_{B-L}$~\cite{Marshak:1979fm,Mohapatra:1980qe}, where $B\!-\!L$ denotes the difference of baryon and lepton number.
However, the identification $U(1) = U(1)_{B-L}$ is not necessary; for the purposes of this Letter, we are in fact able to perform a completely model-independent analysis.
Before we continue, we also note that cosmic strings in field theory (unlike cosmic superstrings) can as well be described by the Abelian Higgs model rather than the Nambu--Goto action, in which case they also lose energy via particle emission~\cite{Vincent:1997cx,Olum:1998ag,Olum:1999sg,Moore:2001px,Hindmarsh:2008dw,Daverio:2015nva,Hindmarsh:2017qff,Matsunami:2019fss}.
The relation between these different modeling approaches is the subject of an ongoing debate in the literature~\cite{Auclair:2019wcv}.
We leave the investigation of the NANOGrav signal from the viewpoint of Abelian-Higgs strings for future work.


In order to compute the GW energy density spectrum, we follow~\cite{Auclair:2019wcv,Cui:2017ufi,Cui:2018rwi} and employ the analytic velocity-dependent one-scale model for cosmic strings~\cite{Martins:1995tg,Martins:1996jp,Martins:2000cs,Sousa:2013aaa,Sousa:2020sxs},
\begin{equation}
\label{eq:OGW}
\Omega_{\rm gw}\left(f\right) = \sum_{k=1}^{\infty} \Omega_{\rm gw}^{(k)}\left(f\right) = \frac{8\pi}{3H_0^2} \left(G\mu\right)^2 f \sum_{k=1}^{\infty} C_k P_k \,,
\end{equation}
where $H_0 \simeq 67\,\textrm{km}/\textrm{s}/\textrm{Mpc}$~\cite{Aghanim:2018eyx} is the present Hubble rate; $G$ is Newton's constant; $\mu$ is the cosmic-string tension (\textit{i.e.}, energy per unit length); $k$ labels the harmonic modes of cosmic-string loops; and $P_k = \Gamma/k^{q}/\zeta\left(q\right)$ is the averaged GW power spectrum.
We assume that $P_k$ is dominated by cusps propagating along cosmic-string loops ($q = 4/3$).
$P_k$ is normalized such that the total emitted power $\Gamma = \sum_k P_k$ agrees with the outcome of numerical simulations, $\Gamma \simeq 50$~\cite{Blanco-Pillado:2013qja,Blanco-Pillado:2017oxo}.
The function $C_k$ in Eq.~\eqref{eq:OGW} is an integral from the onset of the cosmic-string scaling regime, $t_{\rm scl} \ll t_0$, to the present time $t_0$,
\begin{equation}
\label{eq:Ck}
C_k\left(f\right) = \frac{2k}{f^2}\int_{t_{\rm scl}}^{t_0}dt\:\Theta\left(t\right)\left(\frac{a\left(t_{\vphantom{0}}\right)}{a\left(t_0\right)}\right)^5 n\left(\ell_k,t\right) \,.
\end{equation}
Here, $a$ is the cosmic scale factor, which we compute based on the effective numbers of relativistic degrees of freedom $g_\rho$ and $g_s$ tabulated in~\cite{Saikawa:2020swg}.
$t_k$ denotes the time when the loops that contribute to the present-day GW frequency $f$ via their $k^{\rm th}$ harmonic mode were formed,
\begin{equation}
t_k = \frac{\ell_k/t + \Gamma\,G\mu}{\alpha + \Gamma\,G\mu}\,t \,,\quad \ell_k = \frac{2k}{f} \frac{a\left(t_{\vphantom{0}}\right)}{a\left(t_0\right)} \,;
\end{equation}
and $n$ is the number of loops per volume and unit length,
\begin{equation}
n\left(\ell_k,t\right) = \frac{F}{t_k^4}\left(\frac{a\left(t_k\right)}{a\left(t_{\vphantom{0}}\right)}\right)^3 \frac{C_{\rm eff}\left(t_k\right)}{\alpha\left(\alpha + \Gamma\,G\mu\right)} \,.
\end{equation}
$F = 0.1$ is an efficiency factor~\cite{Blanco-Pillado:2013qja,Sanidas:2012ee}; $\alpha = \ell_k/t_k$ characterizes the loop size at the time of formation; and $C_{\rm eff}$ distinguishes between loops formed during radiation ($C_{\rm eff} \simeq 5.4$) and matter ($C_{\rm eff} \simeq 0.39$) domination.
Below, we will simply switch between these discrete values for $C_{\rm eff}$ whenever the dominant form of energy changes.
The $\Theta$ function in Eq.~\eqref{eq:Ck} finally ensures that the time integral only covers physically allowed contributions,
\begin{equation}
\label{eq:theta}
\Theta\left(t\right) = \theta\left(t_0 - t_k\right)\theta\left(t_k-t_{\rm scl}\right)\theta\left(\alpha - \ell_k/t\right) \,.
\end{equation}
In order to evaluate the sum over cosmic-string modes in Eq.~\eqref{eq:OGW}, it is helpful to note that, at large values of $k$,
\begin{equation}
\label{eq:sum}
\sum_{k=m}^n \Omega_{\rm gw}^{(k)}\left(f\right) \approx f^{1-q}\int_{f/n}^{f/m} dx\:x^{q-2}\:\Omega_{\rm gw}^{(1)}\left(x\right) \,,
\end{equation}
which follows from the relation $\Omega_{\rm gw}^{(k)}\left(f\right) = k^{-q}\,\Omega_{\rm gw}^{(1)}\left(f/k\right)$.
Eq.~\eqref{eq:sum} makes it is straightforward to resum a large number of modes.
In our analysis, we include all $k \leq 10^6$.


\begin{figure}
\centering
\includegraphics[width=0.45\textwidth]{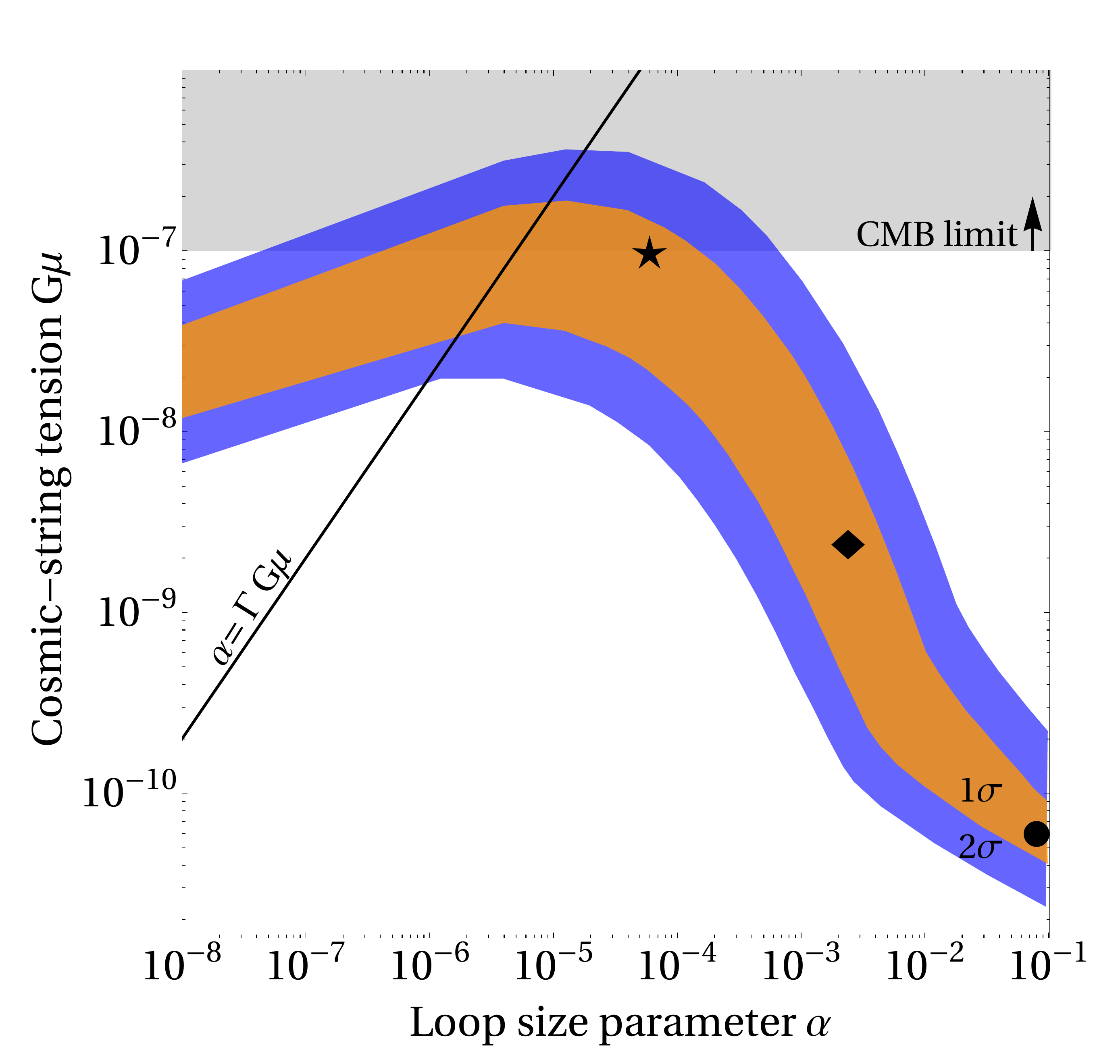}
\caption{NANOGrav $1\,\sigma$ and $2\,\sigma$ posterior contours projected onto the $\alpha$\,--\,$G\mu$ plane.
We do not consider $\alpha$ values larger than $\alpha = 0.1$, which is the maximal value found in simulations.
At $\alpha < 10^{-8}$, we quickly cease to find viable points because $\alpha > \ell_k/t$ in Eq.~\eqref{eq:theta} is no longer satisfied when evaluated in the first NANOGrav frequency bin, \textit{i.e.}, $f \simeq 2\times10^{-9}\,\textrm{Hz}$.
The $\smallblackstar$ benchmark point saturates the CMB limit on $G\mu$ (see Tab.~\ref{tab:bp}).
The diagonal black line labeled $\alpha = \Gamma\,G\mu$ distinguishes between the small-loop and the large-loop regime~\cite{Auclair:2019wcv}.}
\label{fig:scan}
\end{figure}


\smallskip\noindent\textbf{Analysis\,---\,}%
The NANOGrav Collaboration models the pulsar timing-residual cross-power spectrum around a reference frequency $f_{\rm yr} = 1/\textrm{yr}$ by a single power law, $S \propto (f/f_{\rm yr})^{-\gamma}$, with index $-\gamma$.
This cross-power spectrum can be expressed in terms of a characteristic strain
\begin{equation}
h_c\left(f\right) = A \left(\frac{f}{f_{\rm yr}}\right)^{\left(3-\gamma\right)/2} \,,
\end{equation}
which is related to the spectral GW energy density,
\begin{equation}
\label{eq:Ogwfit}
\Omega_{\rm gw}\left(f\right) = \frac{2\pi^2}{3H_0^2}f^2h_c^2\left(f\right) =  \frac{2\pi^2}{3H_0^2}f_{\rm yr}^2\,A^2\left(\frac{f}{f_{\rm yr}}\right)^{5-\gamma} \,.
\end{equation}


By comparing the model prediction in Eq.~\eqref{eq:OGW} with the power-law fit in Eq.~\eqref{eq:Ogwfit}, we are therefore able to determine the amplitude $A$ and the index $\gamma$ as functions of the cosmic-string parameters $G\mu$ and $\alpha$.
The result of this analysis is shown in Fig.~\ref{fig:Agamma}, where we present a scan over $G\mu$ and $\alpha$ in the $\gamma$\,--\,$A$ plane together with the $1\,\sigma$ and $2\,\sigma$ contours determined by NANOGrav.
For all points in the scan, we check that a simple power-law fit provides a good approximation of the actual GW spectrum in the range of frequencies where NANOGrav observes the signal, $f \sim 3\times 10^{-9}\,\textrm{Hz} \cdots 3\times 10^{-8}\,\textrm{Hz}$.
The reason for this is that the NANOGrav signal is confined to a bit less than an order of magnitude in the frequency domain, whereas the cosmic-string-induced GW signal typically only varies on much larger frequency scales.
We checked that fitting the true GW spectrum by a power law introduces an uncertainty in $\gamma$ of at most $\mathcal{O}\left(0.1\right)$.
Remarkably enough, we find that the cosmic-string-induced GW spectrum manages to reproduce the NANOGrav signal across large ranges of the parameters $G\mu$ and $\alpha$.
In particular, we are able to populate the inner region of the $1\,\sigma$ contour.
This would not be possible assuming an SMBHB origin of the signal, which predicts $\gamma = 13/3$. 


\begin{table}

\renewcommand{\arraystretch}{1.35}

\caption{Input ($\alpha$, $G\mu$) and output ($\gamma$, $A$) parameter values for our three benchmark points (see Figs.~\ref{fig:Agamma}, \ref{fig:scan}, and \ref{fig:spectrum}).}

\smallskip
\begin{tabular}{|c||ll|ll|}
\hline
  & \quad$\alpha$ & \quad$G\mu$ & \quad$\gamma$ & \quad$A$ \\
\hline\hline
\:\:$\smallblackstar$\:\:    & \quad $6.0 \times 10^{-5}$ & \quad $1.0\times 10^{-7}$  \quad & \quad $5.13$ & \quad $9.25 \times 10^{-16}$ \quad \\
\:\:$\smallblackdiamond$\:\: & \quad $2.4 \times 10^{-3}$ & \quad $2.4\times 10^{-9}$  \quad & \quad $5.06$ & \quad $9.21 \times 10^{-16}$ \quad \\
\:\:$\smallblackcircle$\:\:  & \quad $1.0 \times 10^{-1}$ & \quad $6.0\times 10^{-11}$ \quad & \quad $4.73$ & \quad $1.28 \times 10^{-15}$ \quad \\
\hline
\end{tabular} 
\label{tab:bp}
\end{table}


In Fig.~\ref{fig:scan}, we provide an alternative visualization of our parameter scan and project the NANOGrav $1\,\sigma$ and $2\,\sigma$ contours onto the $\alpha$\,--\,$G\mu$ parameter plane. 
In this figure, we also indicate the upper bound on the cosmic-string tension, $G\mu \lesssim 10^{-7}$, that follows from the absence of cosmic-string signatures in the CMB~\cite{Ade:2015xua,Charnock:2016nzm,Lizarraga:2016onn}. 
This constraint is derived by considering the effect of long cosmic strings (as opposed to closed cosmic-string loops) on the CMB and is hence independent of $\alpha$.
For $G\mu$ bounds based on the first aLIGO observing run, all of which are weaker than the CMB bound, see Fig. 6 in~\cite{Abbott:2017mem}.
We observe that, despite the strong CMB bound, a significant part of the viable parameter space survives.
Let us now comment on the properties of the GW spectrum in this viable region in dependence of the cosmic-string tension:

\smallskip\noindent{\footnotesize\textbullet}\:$G \mu \lesssim 10^{-11}$:
The signal is weak and cannot be seen.

\smallskip\noindent{\footnotesize\textbullet}\:$10^{-10} \lesssim G \mu \lesssim 10^{-9}$:
The GW signal is strong enough to explain the NANOGrav signal for relatively large $\alpha$ values, $\alpha \sim 0.01\cdots0.1$.
It is interesting to note that this range coincides with the $\alpha$ values that one typically finds in numerical simulations~\cite{Blanco-Pillado:2013qja}.
For smaller $\alpha$ values, the signal decreases and cannot explain the data.

\smallskip\noindent{\footnotesize\textbullet}\:$G \mu \sim 10^{-8}$:
In this regime, a second solution at very small $\alpha$ appears because the peak in the GW spectrum caused by the transition from radiation to
matter domination has the right amplitude to explain the signal. 

\smallskip\noindent{\footnotesize\textbullet}\:$G \mu \sim 10^{-7}$:
Now the GW signal at large $\alpha$ becomes too large. 
One has to go to smaller $\alpha$ values in order to decrease the signal and obtain the right amplitude.
At the same time, the second solution related to the radiation--matter-equality peak at very small $\alpha$ disappears again because the peak starts exceeding the observed signal.

\smallskip\noindent{\footnotesize\textbullet}\:$10^{-6} \lesssim G \mu$:
The GW signal is always too large.

\begin{figure}
\includegraphics[width=0.47\textwidth]{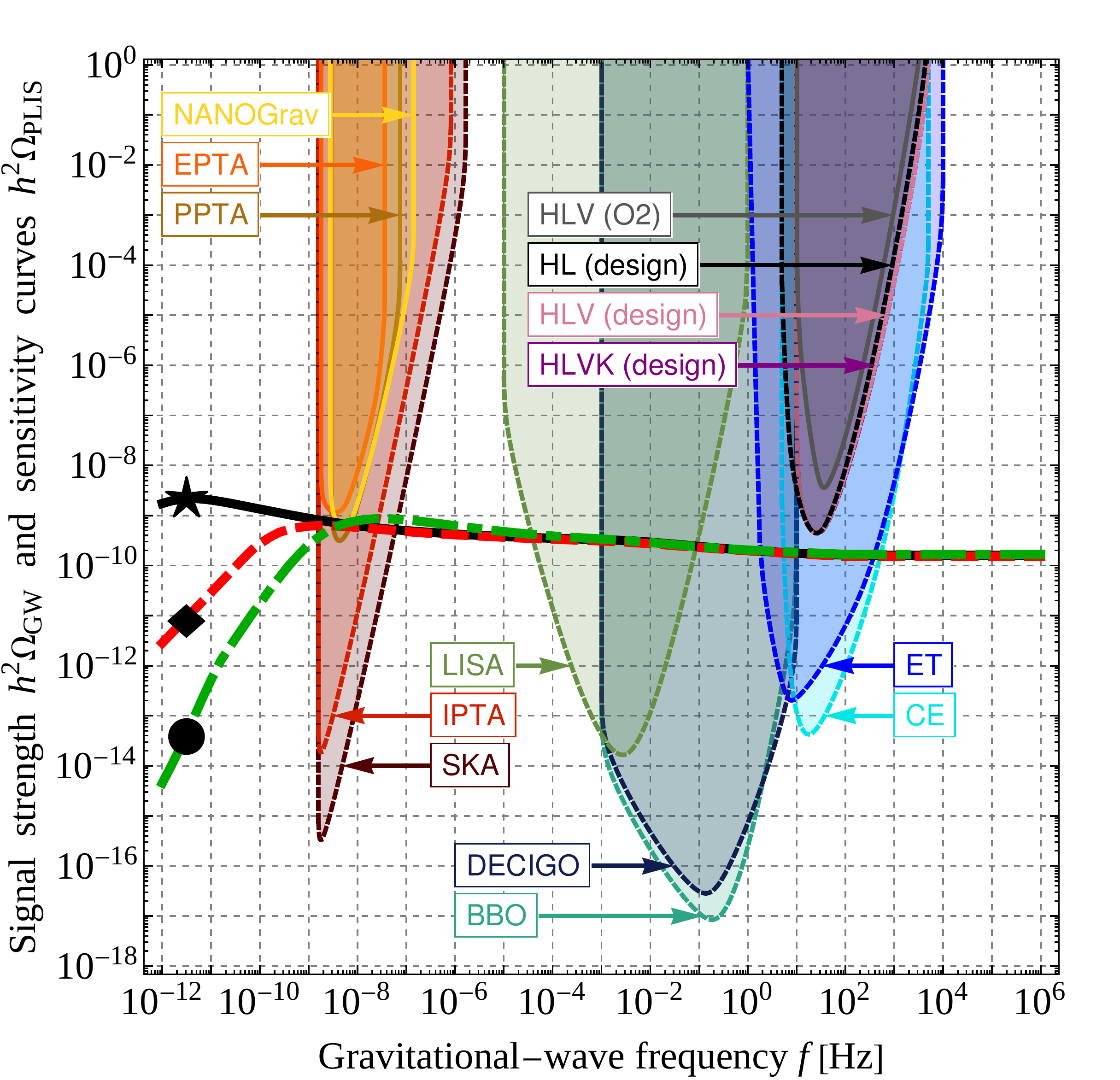}
\caption{GW spectra for the benchmark points ($\smallblackstar$, $\smallblackdiamond$, $\smallblackcircle$) in Tab.~\ref{tab:bp} alongside the power-law-integrated sensitivity curves of various present (solid boundaries) and future (dashed boundaries) GW experiments (see \cite{Schmitz:2020syl} for details).
The EPTA, PPTA, and NANOGrav curves at low frequencies represent the status of PTA constraints on the GW spectrum \textit{prior to the new NANOGrav result}!
Our benchmark spectra therefore illustrate that the NANOGrav signal exceeds previous PTA constraints (see the discussion in the text and \cite{Arzoumanian:2020vkk,Hazboun:2020kzd}).}
\label{fig:spectrum}
\end{figure}

\smallskip\noindent\textbf{Discussion\,---\,}%
If the NANOGrav signal should indeed correspond to a cosmic-string-induced GW spectrum, the implications of this observation would be tremendous.
First of all, we note that the entire viable parameter space in Fig.~\ref{fig:scan} will be probed in future GW experiments.
To show this, we consider the $\smallblackstar$ benchmark point, which saturates the CMB bound on $G\mu$ (see Tab.~\ref{tab:bp}).
We plot the expected GW spectrum for this and the two other benchmark points in Fig.~\ref{fig:spectrum} alongside the \textit{power-law-integrated sensitivity} (PLIS) curves of an array of present and future GW experiments (see \cite{Schmitz:2020syl} for details).
Clearly, the expected spectrum will be within the sensitivity reaches of LISA, DECIGO, BBO, the \textit{Einstein Telescope} (ET), and \textit{Cosmic Explorer} (CE). 
Similarly, future PTA experiments will be able to improve on the current NANOGrav analysis and confirm (or refute) the presence of the signal at increasingly higher significance.
The \textit{LIGO Hanford\,$+$\,LIGO Livingston\,$+$\,Virgo\,$+$\,KAGRA} (HLVK) network, on the other hand, will not be able to detect the signal at a sufficient signal-to-noise ratio.
Next, we note that the height of the flat plateau in the GW spectrum roughly scales as follows in dependence of $\alpha$ and $G\mu$~\cite{Auclair:2019wcv},
\begin{equation}
h^2\Omega_{\rm gw}^{\rm plateau} \simeq 2 \times 10^{-4}\:\bigg(\frac{\bar{\alpha}}{0.1}\bigg)^{1/2}\left(\frac{G\mu}{\Gamma}\right)^{1/2} \,,
\end{equation}
where $\bar{\alpha} = \max\left\{\alpha,9/4\,\Gamma\,G\mu\right\}$.
According to this relation, all viable points in Fig.~\ref{fig:scan} predict a plateau that is at most suppressed by a factor of $\mathcal{O}\big(10^{-3}\big)$ compared to our benchmark spectrum.
As evident from Fig.~\ref{fig:spectrum}, all viable points will therefore be probed in future experiments.


Of course, this statement relies on the assumption of a standard cosmology.
Various nonstandard effects can modify the GW spectrum at high frequencies, including a modified expansion history (\textit{e.g.}, early matter domination), particle production, thermal friction, etc.\ (see \cite{Gouttenoire:2019kij} for an overiew).
Similarly, the cosmic-string network may already form before or during inflation, which would have drastic consequences for the GW signal and constraints on parameter space~\cite{Cui:2019kkd}.
On the one hand, such effects might suppress the GW spectrum, worsening the prospects of detecting the signal at high frequencies.
On the other hand, they might induce nontrivial features in the spectrum that could be probed by space- and ground-based interferometers.
This would open up the possibility to harness the full power of multifrequency GW astronomy and probe physical processes in the early Universe across vast frequency and energy scales.


Finally, we comment on the relation between the cosmic-string tension $G\mu$ and the underlying energy scale $v$ of spontaneous $U(1)$ symmetry breaking~\cite{Bogomolny:1975de,Hindmarsh:1994re},
\begin{equation}
v \sim 10^{16}\,\textrm{GeV} \left(\frac{G\mu}{10^{-7}}\right)^{1/2} \,,
\end{equation}
implying that the NANOGrav signal points to symmetry breaking scales in the range $v\sim 10^{14}\cdots10^{16}\,\textrm{GeV}$. 
This is an exciting result that may indicate a connection between the observed signal and spontaneous symmetry breaking close to the energy scale of grand unification~\cite{King:2020hyd}. 


\smallskip\noindent\textbf{Conclusions\,---\,}%
The NANOGrav Collaboration recently reported strong evidence for a stochastic process across the pulsars in its 12.5-year data set.
In this Letter, we investigated the results of the NANOGrav analysis based on the assumption that this stochastic process corresponds to a primordial SGWB emitted by cosmic strings in the early Universe.
We identified the viable cosmic-string parameter space and argued that the entire viable parameter region will be probed in future GW experiments.
If confirmed in the future, the NANOGrav signal will mark the beginning of a new era in GW astronomy and revolutionize our understanding of the cosmos.


\medskip\noindent\textit{Note added\,---\,}%
Refs.~\cite{Ellis:2020ena,Buchmuller:2020lbh,Samanta:2020cdk,Chigusa:2020rks} also study cosmic strings as the possible origin of the NANOGrav signal.
In fact, the first preprint versions of Ref.~\cite{Ellis:2020ena} and our Letter were released simultaneously.
The authors of Ref.~\cite{Ellis:2020ena} consider a single value of the loop size, $\alpha = 0.1$, for which they find $G\mu \in\left[4,10\right]\times10^{-11}$ at $1\,\sigma$ and $G\mu \in\left[2,30\right]\times10^{-11}$ at $2\,\sigma$.
These bounds are in good agreement with our results, $G\mu \in\left[4,9\right]\times10^{-11}$ at $1\,\sigma$ and $G\mu \in\left[2,22\right]\times10^{-11}$ at $2\,\sigma$.
The remaining discrepancy is well within the uncertainty of our current modeling of cosmic strings.
Besides that, a variety of other scenarios have been studied as possible explanations of the NANOGrav signal since the appearance of our paper, including primordial black holes~\cite{Vaskonen:2020lbd,DeLuca:2020agl,Kohri:2020qqd,Sugiyama:2020roc,Domenech:2020ers}, cosmological phase transitions~\cite{Nakai:2020oit,Addazi:2020zcj,Neronov:2020qrl}, audible axions~\cite{Ratzinger:2020koh,Namba:2020kij}, inflation~\cite{Vagnozzi:2020gtf,Li:2020cjj,Kuroyanagi:2020sfw}, domain walls~\cite{Bian:2020bps,Liu:2020mru}, and a possible violation of the null energy condition~\cite{Tahara:2020fmn}.


\medskip\noindent\textit{Acknowledgments\,---\,}%
We thank Valerie Domcke for helpful comments and Satoshi Shirai for providing us with numerical data on the effective numbers of relativistic degrees of freedom.
We also acknowledge the friendly communication with John Ellis and Marek Lewicki.
This project has received funding from the European Union's Horizon 2020 Research and Innovation Programme under grant agreement number 796961, ``AxiBAU'' (K.\,S.).


\bibliographystyle{JHEP}
\bibliography{arxiv_3}


\end{document}